\documentclass[letterpaper]{article}
\usepackage[latin9]{inputenc}
\usepackage{mathrsfs}
\usepackage{url}
\usepackage{amsmath}
\usepackage{amssymb}
\usepackage{graphicx}
\usepackage{setspace}
\usepackage{esint}
\usepackage[authoryear]{natbib}
\makeatletter

\pdfpageheight\paperheight
\pdfpagewidth\paperwidth

\usepackage{amsfonts}\usepackage{caption}\usepackage{subcaption}

\usepackage{url}\usepackage{multirow}

\usepackage{color}\usepackage{xcolor}

\definecolor{mygreen}{rgb}{0,0.75,0}
\definecolor{mypurple}{rgb}{0.7,0,0.8}

\makeatother

\begin{document}

\title{Fast Out-of-Sample Predictions for Bayesian Hierarchical Models of Latent Health States}

\author{Aaron J Fisher, R Yates Coley, Scott L Zeger}

\date{\today}

\maketitle

\section*{Abstract}

Hierarchical Bayesian models can be especially useful in precision medicine settings, where clinicians are interested in estimating the patient-level latent variables associated with an individual's current health state and its trajectory. Such models are often fit using batch Markov Chain Monte Carlo (MCMC). However, the slow speed of batch MCMC computation makes it difficult to implement in clinical settings, where immediate latent variable estimates are often desired in response to new patient data. In this report, we discuss how importance sampling (IS) can instead be used to obtain fast, in-clinic estimates of patient-level latent variables. We apply IS to the hierarchical model proposed in \citet{Coley2015ProstateSurveillence} for predicting an individual's underlying prostate cancer state. We find that latent variable estimates via IS can typically be obtained in 1-10 seconds per person and have high agreement with estimates coming from longer-running batch MCMC methods. Alternative options for out-of-sample fitting and online updating are also discussed.

\smallskip{}

\section*{Introduction}

Hierarchical Bayesian models provide a natural statistical framework for precision medicine -- allowing estimation of both patient-level latent variables associated with health status trajectory, as well as population-level causal effects of endogenous covariates and exogenous interventions. The use of a Bayesian framework encourages the inclusion of existing medical knowledge at each level. When such models are fit on a training dataset using Markov Chain Monte Carlo (MCMC), latent variable estimates are immediately available for any patient in the training dataset. For example, \citet{Coley2015ProstateSurveillence} use a patient-level latent class to categorize patients in a training dataset as having either indolent or aggressive prostate cancer.

A computational challenge arises though when new patients enter the clinic or when existing patients accrue new measurements. Here, clinicians may wish to give patients fast, in-visit estimates of their latent variables and associated health states. However, batch MCMC estimation approaches that require refitting the entire model can take hours to complete. Additionally, if the model is refit on protected clinical data from multiple sites, then MCMC may require communication between fire-walled servers as the algorithm iterates, further increasing the computation cost.

Instead, sampling algorithms tailored for out-of-sample fitting can be used to get fast latent variable estimates in response to new patient data, while naturally avoiding the issue of server communication. Such algorithms can be based on conditional posteriors \citep{Wu2015}, Gibbs Sampling (GS), Importance Sampling (IS), or Rejection Sampling (RS) \citep{Bishop2006}. In this report, we specifically describe how IS can be used to obtain latent variable estimates for out-of-sample data. We also discuss conceptual parallels between these four types of approaches. We then apply IS to the prostate cancer model proposed by \citet{Coley2015ProstateSurveillence} to get fast latent variable estimates for new, simulated patients. In this case, the IS procedure typically takes approximately 1-10 seconds per patient. This approach can be combined with periodic refitting of the entire model via MCMC, in order to update the posteriors for the population-level parameters \citep{Lee2002}.

This IS approach is related to online (or streaming) learning methods, which aim to continuously update population-level parameters with a constant computational cost over time. We avoid a fully online approach here though, due to additional challenges in online learning. Specifically, our use of IS can be viewed as a 1-step version of a sequential importance sampler (SIS), also known as particle filter. Employing a standard particle filter to update estimates of the population-level parameters would seem to be a natural extension. However, particle filters are known to suffer from the problem of degeneracy, which makes it difficult to estimate posteriors for ``static'' parameters that do not change as more data is acquired (\citet{Kantas2014}, see section II of \citet{Andrieu2005} for an intuitive explanation). This concern applies in our case, as our population-level parameters are assumed to be static. Instead, we combine IS with periodic MCMC \citep{Lee2002} to update the posteriors for all parameters and latent variables at all levels. Note that this is not a fully online method, as the computational cost of MCMC increases as more data is acquired.\footnote{See \citet{Kantas2014} for a recent literature review of particle methods in the context of static parameters.}

Online model fitting has also been explored in the literature on topic modeling for corpuses of texts. Text corpuses are often too large to fit an entire model on at once, making online fitting a more feasible option. \citet{Hoffman2010} propose a online variational Bayes approach for topic modeling. \citet{Canini2009} propose a particle filter approach, in a context where the static parameters can be integrated out. However, in the setting of \citet{Canini2009}, even the best performing online methods were outperformed by batch (non-online) MCMC, and generally did not improve in accuracy as more data was incorporated.

Our specific context within precision medicine \citep{Coley2015ProstateSurveillence} is different than that of topic modeling in that, while the model is complex and contains several layers, the data can be fully stored in memory at once. Thus, while the approach of combining IS with periodic MCMC is not fully online and not feasible for text analysis, it is still a feasible option for the limited sample sizes in our application. Relative to variational Bayes approaches, the formulas required to apply IS are simple to derive and can be easily ported to other applications within precision medicine.

The remainder of this document is organized as follows. In Section \ref{sec:Clinical-Application-=000026}, we give an overview of our motivating data example of latent prostate cancer state estimation. In Section \ref{sec:methods}, we detail our approach for applying IS to hierarchical models. We use an abbreviated notation that can be readily generalized to other precision medicine settings. In Section \ref{sec:Alternative-Out-of-Sample-Poster}, we conceptually compare our IS approach with RS, GS, and conditional posteriors \citep{Wu2015}. However, with the exception of RS, we do not explore the performance of these alternate methods here. In Section \ref{sec:Application}, we apply IS to simulated data, and compare the results to latent variable estimates obtained from batch MCMC.

\section{Clinical Application \& Motivation in Prostate Cancer\label{sec:Clinical-Application-=000026}}

Our application is based on the clinical framework of \citet{Coley2015ProstateSurveillence}, who develop a latent class model to predict underlying prostate cancer state in men participating in an active surveillance program for low risk disease. The latent cancer state is defined as being ``indolent'' or ``aggressive'', corresponding to the Gleason score \citep{Gleason1977,Gleason1992} that would be assigned if a patient's entire prostate were to be removed and analyzed. Gleason scores $<6$ are classified as indolent, and Gleason scores $\geq7$ are classified as aggressive. This latent state is known only for those patients who elect to undergo a prostatectomy while under active surveillance, resulting in a partially latent class problem. For those patients who do not elect to have their prostates removed, the model of \citet{Coley2015ProstateSurveillence} is used to estimate posterior probabilities of latent class membership, or, in other words, the risk of having an aggressive cancer with the potential to metastasize. Latent variable predictions can then be used by clinicians and patients to make decisions about future treatment or biopsies. This prediction tool addresses a pressing need in prostate cancer care as the most common treatments for prostate cancer have a high risk of persistent side effects including erectile dysfunction and urinary incontinence, while prostate biopsies are painful and pose a risk of infection \citep{Chou2011a,Chou2011b}.

The hierarchical model of \citet{Coley2015ProstateSurveillence} includes sub-models for longitudinal prostate specific antigen (PSA) measurements and for longitudinal biopsy results. Both of these sub-models incorporate information about the patient's latent state. More specifically, log-transformed PSA measurements are modeled with a stratified random effects model where the distribution of patient-level random effects depends on latent state. Biopsy results are coded as binary outcomes denoting \textit{grade reclassification} on a biopsy, that is, the biopsied tissue was assigned a Gleason score of 7 or higher. The log-odds of reclassification is modeled with a linear predictor whose value also depends on a patient's latent state, reflecting the imperfect sensitivity and specificity of the biopsy procedure. Each patient's latent class is assumed constant over the surveillance period. As a patient continues in active surveillance, additional PSA and biopsy measurements are accrued and the accuracy of latent class predictions improves. Sub-models are also included for informative observation processes associated with biopsies and prostatectomies.

In our context, the patient-level latent variables refer to the latent classes and the random effects used in the sub-model for PSA. The population-level parameters refer to the coefficients in each sub-model and the variance parameters for the patient-level variables. See \citet{Coley2015ProstateSurveillence} for a full model description.

\section{IS Algorithm for Fast Estimates from New Data\label{sec:methods}}

In this section we detail an IS algorithm that enables rapid estimates of patient-level variables, such as latent classes. This method is meant to be applied to out-of-sample data after MCMC has been applied to get a posterior sample based on current training data. We present the algorithm in a simple, abbreviated notation that is applicable in many clinical settings.

Let the joint posterior based on training data from $n$ patients be denoted as 
\begin{equation}
p(\theta,b_{1:n}|y_{1:n})\propto\prod_{i=1}^{n}[f(y_{i}|b_{i},\theta)g(b_{i}|\theta)]\pi(\theta)\label{eq:posterior_n}
\end{equation}
where $y_{i}$ is the vector of clinical measurements (here, PSA and biopsy measurements) for patient $i$, $y_{1:n}$ is the list of measurements for the first $n$ patients, $b_{i}$ is a vector of latent variables (here, latent class and random effects) for patient $i$, $b_{1:n}$ is a list of latent variables for the first $n$ patients, $\theta$ contains the population-level parameters, $\pi$ is the prior for $\theta$, and $f$ and $g$ are multivariate probability distributions chosen based on the application and context. Estimation of $b_{i}$ is of primary interest in this report.

Let $\mathscr{J}_{n}=\{\theta^{(j)},b_{1:n}^{(j)}\}_{j=1}^{J}$ be a set of $J$ draws from the posterior distribution in Eq \ref{eq:posterior_n} obtained via methods such as MCMC.

\subsection{Core IS Algorithm}

After posterior samples from the joint model $(\mathscr{J}_{n})$ are obtained for current data, importance sampling to update these estimates given new data requires three steps: (1) generating proposal values for the latent variables to be estimated or updated, (2) calculating weights for proposed values, and (3) weighting proposed values to estimate an updated posterior. We first illustrate how this process can be used to quickly estimate latent variables for a new patient and then show how similar calculations can be done to incorporate newly measured data on existing patients in real-time.

For a new patient (indexed by $i=n+1$), prediction of latent variables requires calculating expectations with respect to the posterior distribution based on all $n+1$ patients (i.e. $p(\theta,b_{1:(n+1)}|y_{1:(n+1)})$). While we cannot immediately draw from this distribution, we can evaluate a function that is proportional to its density (based on Eq \ref{eq:posterior_n}). We can also use the posterior distribution based on the first $n$ patients as a proposal distribution (denoted by $q$) from which to generate candidate values of $(\theta,b_{1:(n+1)})$. Let 
\begin{eqnarray}
q(\theta,b_{1:(n+1)}) & := & g(b_{n+1}|\theta)p(\theta,b_{1:n}|y_{1:n})\label{eq:drawing-subj-level-vars}
\end{eqnarray}

Practically, this proposal step is achieved by conditioning on each $\theta^{(j)}$ in $\mathscr{J}_{n}$ and then drawing $b_{n+1}^{(j)}$ from the distribution $g(b_{n+1}^{(j)}|\theta^{(j)})$. This results in the augmented set $\mathscr{J}_{n+1}:=\{\theta^{(j)},b_{1:(n+1)}^{(j)}\}_{j=1}^{J}$. The importance weights $w^{(j)}$ are then proportional to 
\begin{eqnarray}
w^{(j)} & \propto & \frac{p(\theta^{(j)},b_{1:(n+1)}^{(j)}|y_{1:(n+1)})}{q(\theta^{(j)},b_{1:(n+1)}^{(j)})}\nonumber \\
 & \propto & \frac{\prod_{i=1}^{n+1}[f(y_{i}|b_{i}^{(j)},\theta^{(j)})g(b_{i}^{(j)}|\theta^{(j)})]\pi(\theta^{(j)})}{g(b_{n+1}^{(j)}|\theta^{(j)})\prod_{i=1}^{n}[f(y_{i}|b_{i}^{(j)},\theta^{(j)})g(b_{i}|\theta^{(j)})]\pi(\theta^{(j)})}\nonumber \\
 & = & f(y_{n+1}|b_{n+1}^{(j)},\theta^{(j)})\label{eq:importance-weights}
\end{eqnarray}
The final weights $w^{(j)}$ are standardized to sum to 1. The new posterior for $(\theta,b_{1:(n+1)})$ can then be represented as the mixture distribution satisfying $P(\theta=\theta^{(j)},b_{1:(n+1)}=b_{1:(n+1)}^{(j)})=w^{(j)}$. Posterior means for $b_{(n+1)}$ can be calculated as $\sum_{j=1}^{J}w^{(j)}b_{(n+1)}^{(j)}$. 

The approach is similar when we wish to incorporate new measurement data for a patient $k$ who's previous data has already informed the posterior sample $\mathscr{J}_{n}$ (i.e., $k\leq n$). The set $\mathscr{J}_{n}$ already contains proposals $\{b_{k}^{(j)}\}_{j=1}^{J}$ for patient $k$'s latent variable values. Thus, we can use draws from $\mathscr{J}_{n}$ as our proposal distribution $q(\theta^{(j)},b_{1:n}^{(j)})$. Our goal then is to re-weight this set of proposals based on new data. Let $y_{1:n}^{*}$ refer to the data set after incorporating new data on patient $k$, such that $y_{i}^{*}=y_{i}$ if and only if $k\neq i$. The importance weights in Equation \ref{eq:importance-weights} then simplify to 
\begin{eqnarray}
w^{(j)} & \propto & \frac{p(\theta^{(j)},b_{1:n}^{(j)}|y_{1:n}^{*})}{q(\theta^{(j)},b_{1:n}^{(j)})}\nonumber \\
 & \propto & \frac{\prod_{i=1}^{n}[f(\ensuremath{y_{i}^{*}}|b_{i}^{(j)},\theta^{(j)})g(b_{i}^{(j)}|\theta^{(j)})]\pi(\theta^{(j)})}{\prod_{i=1}^{n}[f(\ensuremath{y_{i}}|b_{i}^{(j)},\theta^{(j)})g(b_{i}^{(j)}|\theta^{(j)})]\pi(\theta^{(j)})}\nonumber \\
 & = & \frac{f(\ensuremath{y_{k}^{*}}|b_{k}^{(j)},\theta^{(j)})}{f(\ensuremath{y_{k}}|b_{k}^{(j)},\theta^{(j)})}\label{eq:new-data-patient-ratio}
\end{eqnarray}
If the repeated measures for each patient are independent conditional on $b_{i}$, as is the case in the proposed model from \citet{Coley2015ProstateSurveillence}, then the ratio in Eq \ref{eq:new-data-patient-ratio} reduces to the likelihood of only the new data conditional on $b_{k}^{(j)}$ and $\theta^{(j)}$.

\subsection{Efficient Implementation}

For implementation in clinical practice, proposals for new patients can be generated prior to actually observing new data, so that only weight calculation and  re-weighting of the proposal distribution needs to be done in real-time.

By random chance, some new patients may have data such that very few of the pre-generated, proposed latent variables values receive high weights. This will reduce the effective sample size of the posterior $\left(1/\sum_{j=1}^{J}\left[\left(w^{(j)}\right)^{2}\right]\right)$, which in turn increases the Monte Carlo error of the posterior mean estimates.\footnote{The effective sample size is also known as the effective number of particles.} However, we can use the effective sample size to flag patients who might have high error. When this effective sample size drops below a given threshold (e.g. 1000), we can repeat our procedure with a larger set of pre-generated proposals ($J$). If limited computing is available for MCMC, we can also approximate a larger set of proposals from Eq \ref{eq:drawing-subj-level-vars} by drawing multiple $b_{n+1}$ values for each $\theta^{(j)}$, rather than drawing just one.

\section{Comparison to Alternative Algorithms for Fast Estimates from New Data\label{sec:Alternative-Out-of-Sample-Poster}}

In this section we outline some of the conceptual connections between our IS approach and out-of-sample fitting approaches based on Rejection Sampling (RS), Gibbs Sampling, and conditional posteriors \citep{Wu2015}.

Most directly related to our IS approach, RS can also be applied using the unstandardized weights in Eq \ref{eq:importance-weights}. While RS allows for fewer particles to be stored in memory, we found IS to be more computationally efficient in our scenario (see Section \ref{sub:Results}).

Out-of-sample estimation can also be done using Gibbs Sampling to update only the parameters associated with new patient data (i.e., $b_{n+1}$). One simple implementation is to run separate MCMC chains, each initialized on a different element of $\mathscr{J}_{n}$. Another approximate implementation that combines these chains is to treat the set $\{\theta_{n}^{(j)}\}_{j=1}^{J}$ as fixed and to create a proxy categorical parameter $z$ according to the following hierarchical distribution:

\begin{align*}
\mbox{\textbf{p}} & \sim\text{Dirichlet}(\alpha=\mathbf{1}_{J})\\
z & \sim\text{Categorical}(\mathbf{p})\\
b_{n+1} & \sim g(\theta^{(z)})\\
y_{n+1} & \sim f(b_{n+1},\theta^{(z)})
\end{align*}
where $\mathbf{p}$ is a $J$-length vector of probabilities, $\mathbf{1}{}_{J}$ is a $J$-length vector of ones, and $z$ is a scalar such that $P(z=j)=\mbox{\textbf{p}}_{j}$ for $j=1,2,...J$. The above model can then be fit with traditional Gibbs Sampling, and the resulting posterior estimates for $\mathbf{p}$ are analogous to the weights in Eq \ref{eq:importance-weights}.

Finally, our IS approach functions similarly to the out-of-sample estimation approach of \citet{Wu2015}. Their approach can be generalized to estimate the updated posterior probability that $P(b_{n+1}=x|y_{1:(n+1)})$ using the estimator $\hat{P}(b_{n+1}=x|y_{1:(n+1)}):=\left(\frac{1}{J}\right)\sum_{j=1}^{J}\left\{ \frac{f(y_{n+1}|b_{n+1}=x,\theta^{(j)})g(x|\theta^{(j)})}{\int f(y_{n+1}|b_{n+1}=x',\theta^{(j)})g(x'|\theta^{(j)})dx'}\right\} $. This approach is especially practical when the patient-specific variables $b_{n+1}$ are discrete, and the integral in the denominator can be replaced with a summation. For cases with both continuous and discrete patient-specific variables, the approach can be combined with a proposal generation method based on Eq \ref{eq:drawing-subj-level-vars}. We do not explore the performance of this approach, or of the above Gibbs Sampling approach, in this report.

\section{Application \label{sec:Application}}

We applied the proposed IS approach to simulated data based on the Johns Hopkins Active Surveillance (JHAS) cohort. 1,298 men with very low or low risk prostate cancer diagnoses were enrolled in JHAS from January 1995 to June 2014. Results of all PSA tests and biopsies performed prior to enrollment and during active surveillance were collected. Patients were followed until grade reclassification, elective treatment, or loss to follow-up. Patients still active in the program were administratively censored at the time of data collection for this analysis (October 2014). The Gleason score determination based on pathologic analysis of the entire prostate specimen was also recored for patients who underwent prostatectomy. Details on the dataset are available in \citet{Coley2015ProstateSurveillence}.

Out simulated dataset consisted of 1,000 patients. The model proposed in \citet{Coley2015ProstateSurveillence} was used as the data generating model, with parameter values set equal to their corresponding posterior mean estimates from fitting the model to the JHAS data. Covariates to the model (age and date of diagnosis) were each generated from a normal distribution with mean and variance equal to that observed in JHAS patients. See \citet{Coley2015ProstateSurveillence} for more details on model specification and covariates.

Using this data as our initial sample ($y_{1:n}$), we generate 500,000 draws ($\mathscr{J}_{n}$) from the posterior for the population-level and patient-level variables (see Eq \ref{eq:posterior_n}). Averaging over $\mathscr{J}_{n}$, we estimate the risk of having aggressive cancer for each patient who's latent class is unknown. The task of generating $\mathscr{J}_{n}$ was run across 400 parallel jobs on a x86-based linux cluster, with as many as 200 jobs allowed to run simultaneously. The total elapsed computation time was 33 hours. Within each job, MCMC was implemented using the R2jags software package \citep{R2jags2015}.

We then re-estimate each patient's risk using IS, taking as input only the population-level parameter posteriors from the MCMC step. When generating values for the patient-level variables $b_{i}$, we further increase the diversity of the proposal set by we drawing 10 values from $g(b_{i}|\theta^{(j)})$ for each posterior draw $\theta^{(j)}$, for a total of 5 million proposals. We experimented with approaches of using only 50,000 proposals, using all 5 million proposals, or starting with 50,000 and increasing number of proposals until the effective sample size exceeds 1000. We refer to the first two approaches as ``fixed'' methods and the last approach as ``dynamic''. Within each approach, the final set of proposals were then weighted to obtain IS risk estimates. These simulation steps are meant to approximate the procedure of using IS to get risk estimates for a new patient, under the assumption that any individual patient has only a minor affect on the population-level parameter posteriors. In section \ref{sub:Results}, we assess coherence between IS risk estimates and MCMC risk estimates.

\subsection{Results\label{sub:Results}}

\begin{figure}
\centering{}\includegraphics[width=0.55\textwidth]{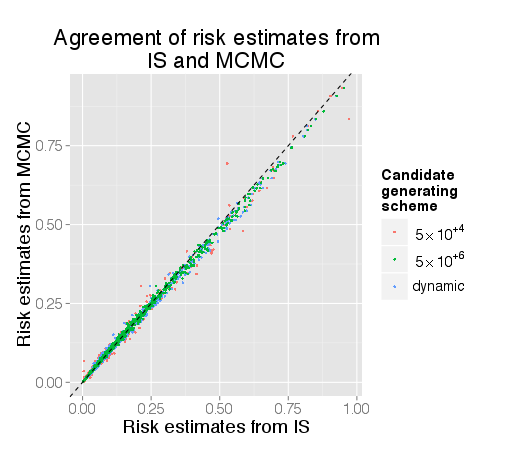} \caption{Agreement between IS and MCMC estimates for the posterior predictions of aggressive prostate cancer state in a new patient. Point color represents the number of candidate points used -- 50,000, 5 million, or dynamic. The dashed line indicates the axis of equality (i.e., perfect agreement).\label{fig:jags-vs-pf} }
\end{figure}

We find a high degree of coherence between the estimated posterior probability of aggressive cancer from IS and from MCMC, as shown in Figure \ref{fig:jags-vs-pf}. With the dynamic proposal approach, the root mean square of the difference (rMSD) between these two sets of risk estimates was 1.3\% (on the probability scale, from 0\% to 100\%). The maximum absolute difference was 5.9\%, with 99\% of patients having a difference less than 4.5\%. Of the approaches using a fixed number of proposals, the corresponding maximum and 99\% quantile of differences were 16.6\% and 5.6\% for 50,000 proposals and 4.8\% and 3.5\% for 5 million proposal. We also considered a rejection sampling approach using the unstandardized weights in Eq \ref{eq:importance-weights} but found the results to have a greater deviation from MCMC estimates (rMSD = 1.6\% for the dynamic approach). 

Figure \ref{eq:importance-weights} illustrates the roughly inverse relationship between the effective sample size used for IS and the difference between IS and MCMC risk estimates.\footnote{It is worth noting, however, that the differences did not tend to decrease at a rate proportional to the square root of the effective sample size.} Here, all three approaches performed similarly, with the fixed 50,000-proposal approach being the fastest. Computation time per patient ranged from 1.5-13.5 minutes, 1.4 seconds - 5.9 minutes, and 1.4-2.8 seconds for the 5-million-proposal, dynamic, and 50,000-proposal approaches, respectively. Interquartile ranges for the per-patient calculation times of three methods were 3.1-3.9 minutes, 2.3-4.6 seconds, and 2.3-2.5 seconds.

These findings suggest that the proposed IS algorithm can be an appropriate substitute for full MCMC runs in order to provide real-time updates in a clinical setting.

\begin{figure}
\centering{}\includegraphics[width=0.55\textwidth]{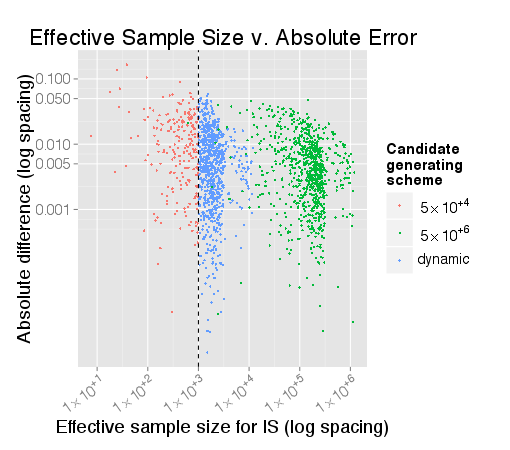} \caption{Difference between IS and MCMC risk estimates as a function of effective sample size for IS. Point color represents the number of candidate points used -- 50,000, 5 million, or dynamic. The dotted vertical line shows the threshold used for dynamic proposal generation at an effective sample size of 1,000. Both axes are shown with log-scale spacing. \label{fig:effective-ss-v-deviation}}
\end{figure}

\section{Discussion}

The joint model of \citet{Coley2015ProstateSurveillence} is among a growing number of statistical models for making individualized health predictions and recommendations. Development of such \textit{\emph{precision medicine}} methods must occur within a framework for clinical implementation. Specifically, concerns about convenience, security, and effective communication must be addressed alongside statistical considerations. In this technical report, we present a fast implementation of the of latent health state model proposed in \citet{Coley2015ProstateSurveillence}, using importance sampling to generate in-clinic predictions. This approach informs decision-making by enabling doctors and patients to access updated predictions in real-time in a clinical setting.

\section*{Supplemental Code}

Code for simulating data, obtaining IS estimates, and comparing the results against MCMC estimates is available at: \url{https://github.com/aaronjfisher/in-clinic-updates-PSA}

\bibliographystyle{apalike}
\bibliography{inhealth-bib}

\begin{thebibliography}{}

\bibitem[Andrieu et~al., 2005]{Andrieu2005}
Andrieu, C., Doucet, A., and Tadic, V.~B. (2005).
\newblock On-line parameter estimation in general state-space models.
\newblock In {\em Decision and Control, 2005 and 2005 European Control
  Conference. CDC-ECC'05. 44th IEEE Conference on}, pages 332--337. IEEE.

\bibitem[Bishop et~al., 2006]{Bishop2006}
Bishop, C.~M. et~al. (2006).
\newblock {\em Pattern recognition and machine learning}, volume~4.
\newblock springer New York.

\bibitem[Canini et~al., 2009]{Canini2009}
Canini, K.~R., Shi, L., and Griffiths, T.~L. (2009).
\newblock Online inference of topics with latent dirichlet allocation.
\newblock In {\em International conference on artificial intelligence and
  statistics}, pages 65--72.

\bibitem[Chou et~al., 2011a]{Chou2011a}
Chou, R., Croswell, J.~M., Tracy, D., Bougatsos, C., Blazina, I., Fu, R.,
  Gleitsmann, K., Koenig, H.~C., Lam, C., Maltz, A., Rugge, J.~B., and Lin, K.
  (2011a).
\newblock {Screening for prostate cancer: a review of the evidence for the U.S.
  Preventive Services Task Force}.
\newblock {\em {Annals of Internal Medicine}}, 155:762--771.

\bibitem[Chou et~al., 2011b]{Chou2011b}
Chou, R., Dana, T., Bougatsos, C., Fu, R., Blazina, I., Gleitsmann, K., and
  Rugge, J.~B. (2011b).
\newblock {Treatments for Localized Prostate Cancer: Systematic Review to
  Update the 2002 U.S. Preventive Services Task Force}.
\newblock Evidence Synthesis No. 91. ARHQ Publication No. 12-0516-EF-2.
  Rockville, MD: Agency for Healthcare Research and Quality.

\bibitem[Coley et~al., 2015]{Coley2015ProstateSurveillence}
Coley, R.~Y., Fisher, A.~J., Mamawala, M., Carter, H.~B., Pienta, K.~J., Zeger,
  and L, S. (2015).
\newblock Bayesian joint hierarchical model for prediction of latent health
  states with application to active surveillance of prostate cancer.
\newblock (\url{http://arxiv.org/abs/1508.07511}).

\bibitem[Gleason, 1977]{Gleason1977}
Gleason, D. (1977).
\newblock {The Veteran's Administration Cooperative Urologic Research Group:
  Histologic grading and clinical staging of prostatic carcinoma}.
\newblock In Tannenbaum, M., editor, {\em {Urologic Pathology: The Prostate}},
  pages 171--198. Lea and Febiger, Philadelphia.

\bibitem[Gleason, 1992]{Gleason1992}
Gleason, D.~F. (1992).
\newblock Histologic grading of prostate cancer: a perspective.
\newblock {\em Human pathology}, 23(3):273--279.

\bibitem[Hoffman et~al., 2010]{Hoffman2010}
Hoffman, M., Bach, F.~R., and Blei, D.~M. (2010).
\newblock Online learning for latent dirichlet allocation.
\newblock In {\em Advances in neural information processing systems}, pages
  856--864.

\bibitem[Kantas et~al., 2014]{Kantas2014}
Kantas, N., Doucet, A., Singh, S.~S., Maciejowski, J.~M., and Chopin, N.
  (2014).
\newblock On particle methods for parameter estimation in state-space models.
\newblock {\em arXiv preprint arXiv:1412.8695}.

\bibitem[Lee and Chia, 2002]{Lee2002}
Lee, D.~S. and Chia, N.~K. (2002).
\newblock A particle algorithm for sequential bayesian parameter estimation and
  model selection.
\newblock {\em Signal Processing, IEEE Transactions on}, 50(2):326--336.

\bibitem[Su and Yajima, 2015]{R2jags2015}
Su, Y.-S. and Yajima, M. (2015).
\newblock {\em R2jags: A Package for Running jags from R}.
\newblock R package version 0.05-01.

\bibitem[Wu et~al., 2015]{Wu2015}
Wu, Z., Deloria-Knoll, M., Hammitt, L.~L., and Zeger, S.~L. (2015).
\newblock Partially latent class models for case--control studies of childhood
  pneumonia aetiology.
\newblock {\em Journal of the Royal Statistical Society: Series C (Applied
  Statistics)}.

\end{thebibliography}

\end{document}